\documentstyle[floats,multicol,aps,epsf,prb]{revtex}

\newcommand \beq  {\begin{equation}}
\newcommand \eeq  {\end{equation}}
\newcommand \bea {\begin{eqnarray} }
\newcommand \eea {\end{eqnarray}}

\begin{document}
\draft
\twocolumn[\hsize\textwidth\columnwidth\hsize\csname @twocolumnfalse\endcsname
\title{Superconducting Quantum Critical Point}
\author{R. Ramazashvili and  P. Coleman}
\address{
Serin Laboratory, Rutgers University, P.O. Box 849,
Piscataway, NJ 08855-0849, USA.}
\maketitle
\date{\today}
\maketitle
\begin{abstract}
We study the properties of a quantum critical
point which develops in a BCS superconductor when pair-breaking
suppresses the transition temperature to zero. The pair
fluctuations are characterized by a dynamical critical exponent
$z=2$. Except for very low temperatures, anomalous contribution 
to the conductivity is proportional to $\sqrt T$ in three dimensions,
but to $1/T$ in two dimensions. At lowest temperatures, the conductivity 
correction varies as $T^{1/4}$ in three dimensions, and as 
$\ln{(1/T)}$ in two. 
\end{abstract}
\vskip 0.2 truein
\pacs{PACS numbers: 71.10.Hf, 71.27.+a, 74, 74.20.Fg}
\vskip2pc]
The possibility of quantum critical behavior in itinerant magnets
has attracted great attention in recent years,
in part, because quantum criticality affords the possibility
of a controlled study of non-Fermi liquid
behavior.\cite{lonzarich} At a quantum critical
point, order parameter fluctuations develop an infinite correlation
range in both space {\em and} time.\cite{hertz,millis} 
The coupling between these
fluctuations and conducting electron fluid is able, under certain
circumstances, to eliminate the formation of well-defined
quasiparticles in the electron liquid, giving rise to a new
kind of metallic behavior.

In this paper we discuss the possibility of quantum critical
behavior in superconductors. A quantum critical point implies
a finite value of the electron interaction strengths.  
At first sight, this would appear to rule out the possibility 
of a superconducting quantum critical point, for conventional 
superconductivity develops for an arbitrarily small pair 
interaction. If the transition temperature is driven to zero in
a pure BCS superconductor, the pairing interaction and the pair
fluctuations are completely eliminated. Fortunately, this is not
the case in the presence of pair-breaking, which cuts off the
logarithmic singularity in the pair susceptibility, requiring
that the pairing interaction reach a critical strength before
superconductivity develops.  In this paper we characterize
the quantum critical behavior which develops at
this special point.  Our results can be tested experimentally
on conventional superconductors, such as $Ce$-doped $La$. \cite{maple} 
They may also provide a useful diagnostic tool for the understanding 
of unconventional, e.g. heavy fermion superconductors.

\begin{figure}
\epsfxsize=3.2 truein
\epsfbox{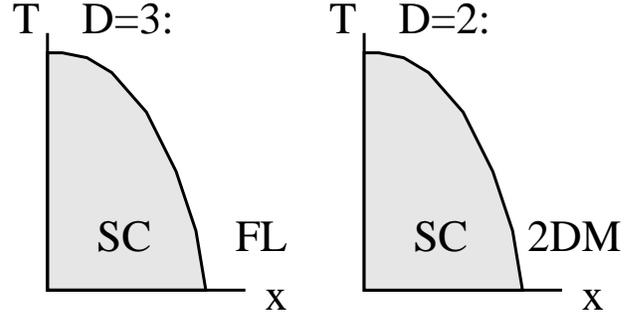}
\caption{Schematic phase diagram in 
three ($D=3$) and in two ($D=2$) spatial dimensions. Shaded regions
correspond to the superconducting (SC) phase. 
The normal phase 
corresponds to the Fermi liquid (FL) in three dimensions, 
and to a two-dimensional disordered metal (2DM) in two.
}
\end{figure}

We begin with writing down the effective action of the pair field $\Delta$
in the vicinity of a quantum critical point:
\beq
S_{eff}[\Delta]=T\sum_{\nu,k} \Delta^{\dagger}\chi^{-1}\Delta 
+ \frac{N(0) T^3}{T_{0}^2} \sum_{1,2,3} 
\Delta^{\dagger}_1\Delta^{\dagger}_2\Delta_3\Delta_4,
\eeq
\beq
\chi^{-1}=N(0)\left[\frac{\delta x}{x_c} 
+ \left[\frac{T}{T_{0}}\right]^2 +
\frac{|\nu|}{T_{0}} + \left[\frac{v_F q}{T_{0}}\right]^2
\right],
\eeq
which is valid for $T, vq$, $|\nu| << T_{0}$. Here  $T$ is temperature,
N(0) is the density of states at the Fermi surface, $\nu$ is a Matsubara 
frequency, $q$ is momentum, $v_F$ is the Fermi velocity, and $\delta x$
is the deviation of the control parameter $x$ from its critical value $x_c$,
where the transition temperature turns into zero.

We obtained this action by repeating the Abrikosov-Gorkov
calculation \cite{abrikosov} for an $s$-wave BCS superconductor 
doped with magnetic impurities. Apart from the original calculation
\cite{abrikosov}, the frequency and momentum dependence of the 
disorder-averaged pair propagator was kept and at the end 
both the pair field and the momentum were rescaled to give (1)-(2). 

The quantity $T_{0}$ is the only characteristic energy scale 
of the effective action (1)-(2). In a BCS superconductor 
it is of the order of the pair-breaking rate, which at the critical point 
is of the order of the transition temperature in the clean system. 
Both quantities are much less than the Fermi energy $\epsilon_F$. 
However, in a strongly correlated material, $T_{0}$ may, 
in principle, be of the order of $\epsilon_F$. 

These two limiting cases correspond to different physics. 
In the BCS case, the quartic term in (1) can be neglected at all 
experimentally relevant temperatures $T$ well below $T_{0}$. By contrast,
if $T_{0} \sim \epsilon_F$, the feedback of the quartic term 
dominates at $T << T_{0}$. In the latter case, one has to regard $T_{0}$ 
as a phenomenological parameter, resulting from a strong coupling, 
or non-BCS pairing mechanism.

\begin{table}
\protect\caption{ Order of magnitude and leading temperature dependences 
of the imaginary part of the electron self-energy $\Sigma(\omega; p=p_F; T=0)$
due to scattering by the fluctuations of $\Delta$, the specific heat 
coefficient correction $\delta C(T)/T$ and the conductivity correction 
$\delta \sigma(T)$ in the 
weak coupling (BCS) limit 
($T_{0} << \epsilon_F$)
and in the strong coupling limit ($T_{0} \sim \epsilon_F$). 
The value of $\sigma_0$ corresponds to the residual normal state 
conductivity.}
\begin{center}
\begin{tabular}{cccc}
  & Electron    & Specific Heat & Conductivity  \\
  & Self-Energy & Coefficient   & Correction    \\
  & ${\displaystyle Im \Sigma(\omega)}$  & 
${\displaystyle \delta C(T)/T}$ & ${\displaystyle \delta \sigma(T)}$ \\
\tableline
$T_{0} << \epsilon_F,$ & & & \\
$d=3 :$
& ${\displaystyle \frac{\left(T_{0}\omega\right)^{3/2}} 
{\epsilon_F^2}}$ 
&${\displaystyle \frac{T_{0}^{3/2}T^{1/2}}
{\epsilon_{F}^3}} $ &
${\displaystyle \sigma_0 \frac{T_{0}^{3/2}T^{1/2}} {\epsilon_{F}^2}}$ \\
 & & & \\ 
\tableline
$T_{0} \sim \epsilon_F,$ & & & \\
$d=3 :$ & ${\displaystyle \frac{\omega^{3/2}}{\epsilon_F^{1/2}}}$ 
 &${\displaystyle \frac{T^{1/2}}{\epsilon_F^{3/2}}}$& 
 ${\displaystyle \sigma_0 \left(\frac{T}{\epsilon_F}\right)^{1/4}}$ \\
  & & & \\
 \tableline
$T_{0}<< \epsilon_F,$ & & & \\
$d=2 : $ & ${\displaystyle \frac{\omega T_{0}}{\epsilon_F}}$ & 
 ${\displaystyle \frac{T_{0}}{\epsilon_{F}^2}
\ln\left(\frac{T_{0}}{T}\right)}$ &
${\displaystyle \sigma_0 \frac{T_{0}^2}{T \epsilon_F}}$ \\
 & & & \\
\tableline
$T_{0} \sim \epsilon_F,$& & & \\
$d=2 : $ & ${\displaystyle \omega}$ &  
${\displaystyle \frac{1}{\epsilon_{F}}
\ln\left(\frac{\epsilon_{F}}{T}\right)  }$& 
${\displaystyle 
\sigma_0 \ln\left(\frac{\epsilon_{F}}{T}\right)}$ \\
\end{tabular}
\end{center}
\end{table}

With (1)-(2) at hand, one can calculate various 
thermodynamic and transport properties of interest.
Table I presents the results for the zero-temperature quasiparticle
decay rate due to scattering by the $\Delta$ field, and for the leading 
corrections to low-temperature thermodynamics and transport 
at $T_c=0$. 
In the BCS limit, corrections to the specific heat 
coefficient come from Gaussian fluctuations of the pair field.\cite{tsvelik} 
The correction for a strong coupling case was found in \cite{millis} using
renormalization group methods. 
The quasiparticle decay rate due to scattering off pair fluctuations
is estimated by the diagram on Fig.2 (a). It is essential
for the calculation of conductivity and of the quasiparticle decay rate, 
that the electron vertex corrections are not singular, 
since the pair-breaking makes the lifetime of a Cooper pair finite. 
The leading conductivity correction is given 
by the Aslamazov-Larkin diagram \cite{alar} shown on Fig.2(b), which can 
be regarded as conductivity of particles with the inverse propagator 
given by (2), at $\delta x =0$. 

The calculation was done by renormalization group analysis
of the expression for the Aslamazov-Larkin correction 
on Fig.2 (b). After transforming the Matsubara sum into a contour 
integral and going to dimensionless variables, it reads \cite{arhilar}:
\beq
\Delta\sigma=\int^1 q^2 d^Dq \int^1 \frac{dz}{T} \frac{1}{\sinh^{2}
\frac{z}{2T}} \left[ Im \chi(q,z+i0) \right]^{2},
\eeq
where both the energy and the momentum cut-off have been set to unity.
When renormalizing, we will follow \cite{millis}: first integrate out 
a thin outer shell in the momentum space between $1$ and $1 - 1/b$. 
Then rescale the momentum ($q \rightarrow q b$) to restore the cut-off, 
then rescale the energy ($z \rightarrow zb^2$), the mass term 
and the temperature ($T \rightarrow Tb^2$) and, finally, integrate out 
an energy shell to restore the energy cut-off. 
At each step the quartic interaction induces corrections to the mass term 
and to itself (see \cite{millis} for details of the 
renormalization group equations). 

As a result, one arrives at the following 
transformation law for the Aslamazov-Larkin correction: 
$$\Delta\sigma[J]=b^{2-D}\Delta\sigma[J'] + \ln b f[J],$$
where $J$ denotes the mass term, the temperature and 
the quartic coefficient, $J'$ denotes their renormalized values
and $D$ is the dimensionality of the sample. 
The precise form of $f[J]$ can be easily obtained using the above
renormalization procedure. However, only two features of $f[J]$ 
are important: 
(a) as long as the running value of $T$ is smaller 
than the cut-off, $f[J]$ has rather weak dependence on its arguments; 
this corresponds to the quantum renormalization region; 
(b) in the classical renormalization region, 
when $T$ exceeds the cut-off, $f[J]$ is proportional to $T$. 
Thus $\Delta\sigma$ can be written as 
\beq
\Delta\sigma=\int_{0}^{\ln b^*} dx f[J(e^{x})] e^{(2-D)x},
\eeq
where $b^*$ is the value of the rescaling factor at which the mass term 
reaches the cut-off and the scaling process stops. It is of the order 
of $T^{-3/4}$ in three dimensions and of the order 
of $\sqrt{\ln{(1/T)}/T}$ in two.\cite{millis} To evaluate (4),
one also needs the value of $b_1$ such that $T(b_1) \sim 1$, which is 
$b_1 \sim T^{-1/2}$ independently of the dimensionality. With these 
prerequisites, the answers in Table I follow as soon as one neglects 
the dependence of $f[J(e^{x})]$ on all the couplings except temperature:
$$ f[J(e^{x})]=f[Te^{2x}].$$ 

In three dimensions, the same result can be obtained just by 
replacing the ``bare" $T^2$ mass term in (2) by its renormalized
value $T^{3/2}$, and then calculating the conductivity correction 
as per (3). 

Scaling 
analysis also allows to show 
that the vertex corrections are negligible. Their inclusion 
reduces to putting into the diagram of Fig.2 (b) extra bubbles 
such as one on Fig.2 (c). Each bubble contributes two Green's functions 
plus an integral over the energy and momentum. After an infinitesimal 
scaling transformation, this gives an extra factor of $b^{2-D}$. 
Thus, in three dimensions, vertex corrections are irrelevant. 
In two dimensions, they appear marginal yet do not introduce any 
extra corrections. This can be established by using the Ward identity
and writing the renormalization 
equation directly for the current vertex ( see Fig.3), and then inserting 
the solution into the corresponding expression for the conductivity. 

We would like to discuss theoretical and experimental 
implications of the results.
As envisaged by Hertz \cite{hertz}, 
under quite general circumstances 
the superconducting quantum critical point falls into the $z=2$ 
universality class. Although in principle different behavior 
cannot be ruled out, it appears to require additional tuning of parameters, 
as well as rather unusual features of the pairing phenomenon itself,
such as the gap vanishing at the entire Fermi surface.\cite{us} 

In a weakly interacting disordered metal, $T_{0}$ in (2) 
is of the order of the 
impurity
scattering rate. In this case 
anomalous corrections due to quantum criticality 
in the Landau-Ginzburg regime are of the order of the weak localization 
corrections.\cite{altar} 
Therefore, 
in two dimensions 
our results are fully consistent only as long 
as all corrections are small and additive. 
Moreover, in two dimensions, the electron-electron interaction 
is known to lead to the linear temperature dependence 
of the quasiparticle decay rate \cite{altar,time}, regardless 
of closeness to the quantum critical point. Thus, for a weakly 
interacting two-dimensional system, the entire normal 
region corresponds to the two-dimensional disordered 
metallic regime (see Fig.1) with the quasiparticle decay rate 
proportional to the quasiparticle energy. 

However, 
we expect that 
in a strongly correlated system
with suppressed weak localization effects, superconducting quantum
critical point still falls into the $z=2$ universality class and 
Table I may describe the actual state of affairs 
for all dimensionalities. 
In this case, Fermi liquid turns marginal only at $T_c=0$
and crosses over to the normal Fermi liquid behavior 
as the system is driven away from the quantum critical point 
into the metallic phase.

Finally, we should like to comment on the diagnostic opportunities,
furnished by measurements at a superconducting 
quantum critical point.
In light of the discussion in the beginning 
of the paper, one is led to conclude that in a {\em clean} time reversal 
invariant system, observation of singular behavior at the superconducting 
quantum critical point would mark a {\em very} peculiar phenomenon, 
as in a clean BCS superconductor suppression 
of $T_c$ completely eliminates pair fluctuations. 
Since any sample contains impurities, in reality the above conclusion 
refers to temperatures above the elastic scattering rate:
$T > \tau^{-1}$. The latter can be extracted from the residual 
resistivity measurements of the sample. 

To summarize, we studied the properties of a superconductor near a
quantum critical point driven by pair-breaking disorder. 
Superconducting fluctuations 
are characterized by a dynamical critical 
exponent $z=2$. In a BCS superconductor at experimentally accessible 
low temperatures, the singular contribution to the conductivity 
is proportional to $\sqrt T$ in three dimensions, but to $1/T$ 
in two dimensions. In a superconductor with the characteristic 
energy scale of the order of $\epsilon_F$ at the quantum critical point,
the contribution to the conductivity varies as $T^{1/4}$ in three dimensions,
and as $\ln{(1/T)}$ in two. 

We are indebted to E. Abrahams, I. Aleiner, L. Glazman, G. Kotliar, 
A. Larkin, A. Millis, M. Stephen and A. Tsvelik for discussions 
related to this work, which was supported by the National Science 
Foundation under Grant number DMR-96-14999.
We would like to thank the Theoretical Physics group at Oxford 
for the kind hospitality during the period when this work was 
completed. 

\begin{figure}
\epsfxsize=3.2 truein
\epsfbox{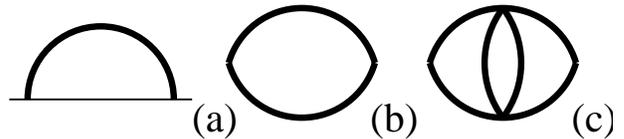}
\caption{ (a) Graph for the quasiparticle self-energy due to scattering
by pair fluctuations. Solid lines denote the pair propagator; thin line
denotes the quasiparticle Green function. (b-c) Graphs for the conductivity 
correction: (b) the Aslamazov-Larkin graph (c) an example of a graph 
with vertex corrections. Scaling dimension of all such graphs 
is an integer multiple of $(2-D)$.}
\end{figure}

\begin{figure}
\epsfxsize=3.2 truein
\epsfbox{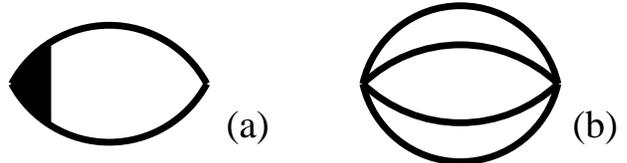}
\caption{ (a) Graph for the conductivity, including the current vertex,
denoted by the dark triangle.
(b) contribution to the free energy in the second order in 
the quartic term, which gives rise to renormalization of the current 
vertex.} 
\end{figure}


\begin{references}
\bibitem{lonzarich} see J. Phys. Cond. Mat. {\bf 8}, (1996), for exhaustive
references. 
\bibitem{hertz} John A. Hertz, Phys. Rev. {\bf B 14}, 1165 (1976).
\bibitem{millis} A. J. Millis, Phys. Rev. {\bf B 48}, 7183 (1993).
\bibitem{maple} M. B. Maple {\em et al.}, Phys. Rev. Lett. 
{\bf 23}, 1375 (1969).
\bibitem{abrikosov} A. A. Abrikosov, L. P. Gor'kov, Sov. Phys. JETP 
{\bf 12}, 1243 (1961).
\bibitem{tsvelik} see, e.g., A. M. Tsvelik, {\it Quantum Field Theory 
in Condensed Matter Physics}, Cambridge University Press, 1995.
\bibitem{alar}  L. G. Aslamazov, A. I. Larkin, Sov. Phys. Solid State 
{\bf 10}, 1104 (1968).
\bibitem{arhilar}  A. G. Aronov, S. Hikami, A. I. Larkin, 
Phys. Rev. {\bf B 51}, 3880 (1995). 
\bibitem{us} R. Ramazashvili, Phys. Rev. {\bf B 56}, 5518 (1997), 
and references therein. 
\bibitem{altar} B. L. Altshuler, A. G. Aronov, in {\it Electron-Electron
Interactions in Disordered Systems}, editors A. L. Efros and M. Pollak
(Elsevier Science Publishers B. V., 1985).
\bibitem{time} E. Abrahams, P. W. Anderson, P. A. Lee, 
and T. V. Ramakrishnan, Phys. Rev. {\bf B 24}, 6783 (1981); 
\end{references}
\end{document}